\documentclass[%
 reprint,
 amsmath,amssymb,
 aps,
]{revtex4-2}

\usepackage{graphicx}
\usepackage{dcolumn}
\usepackage{bm}
\usepackage{comment}
\usepackage{textcomp}
\usepackage{amsmath}
\usepackage{mathrsfs}

\usepackage{amssymb}
\usepackage{units}
\usepackage{ulem}
\usepackage{xcolor}
\usepackage[colorlinks,bookmarks=false,citecolor=blue,linkcolor=blue,urlcolor=blue]{hyperref}
\usepackage{soul}

\begin{document}

\title{Beyond-mean-field 
corrections to the blueshift of a driven-dissipative exciton-polariton condensate}%

\author{Félix Helluin$^1$,  Léonie Canet$^{1,2}$, Anna Minguzzi$^1$}%
\affiliation{$^1$ Université Grenoble Alpes, CNRS, LPMMC, 38000 Grenoble France, $^2$ Institut Universitaire de France, 5 rue Descartes, 75005 Paris}


\begin{abstract}
In the absence of vortices or phase slips, the phase dynamics of exciton-polariton condensates was shown to map onto the Kardar-Parisi-Zhang (KPZ) equation, which describes the stochastic growth of a classical interface. This implies that the coherence of such non-equilibrium quasi-condensates decays in space and time following stretched exponentials, characterized by KPZ universal critical exponents. In this work, we  focus on the time evolution of the average phase of a one-dimensional exciton-polariton condensate in the KPZ regime and determine the frequency of its evolution, which is given by the blueshift, i.e. the non-equilibrium analog of the chemical potential. We determine the stochastic corrections to the blueshift within Bogoliubov linearized theory and find that while this correction physically originates from short scale effects, and depends both on density and phase fluctuations, it  can still be related to the effective large-scale KPZ parameters.  Using numerical simulations of the full dynamics, we investigate the dependence of these blueshift corrections on both noise and interaction strength, and compare the results to the Bogoliubov prediction. Our finding contributes  both to the close comparison between equilibrium and non-equilibrium condensates, and to the theoretical understanding of the KPZ mapping.
 \end{abstract}

\maketitle

\section{\label{sec:Intro} Introduction}

Exciton-polaritons (EPs) are bosonic quasi-particles that arise from the strong coupling between light and matter \cite{hopfield1958}. They are typically formed in a quantum well embedded in an optical microcavity, from the interaction between quantum well excitons and cavity photons \cite{savonaPolariton, baboux2018}. The system undergoes unavoidable losses due to the leakage of cavity photons, which  have to be compensated by a continuous laser drive in order to sustain a non-equilibrium steady 
state of EPs. Remarkably, a  Bose-Einstein condensate (BEC) can form in these driven-dissipative conditions, as was first experimentally demonstrated in \cite{nature1stPolaritonBEC}, and yields  a wide range of novel physical phenomena \cite{carusottoQFL, bloch2022ReviewOOEBEC}. Among them, recent studies established a connection with the Kardar-Parisi-Zhang (KPZ) universality class \cite{altmanPRX2015KPZmapping,Ji2015, Diehlspacetimevortex, squizzato2018KPZsubclasses, deligiannis2021KPZsubclasses, deligiannis2022KPZ2D, amelio2020theory}, known in classical statistical physics to describe the scale invariant properties of kinetically roughening interfaces and characterized by a  superdiffusive universal exponent \cite{KPZ1986}. This connection was first observed in experiments for a one-dimensional (1D) EP condensate in Ref.~\cite{Bloch&co2021}, which reported the expected stretched exponential decay of the first order correlation function, settling a major difference with coherence properties of 1D equilibrium quasi-condensates \cite{haldane,giamarchi-book}, which display power-law (or simple exponential) decay at zero (or finite temperature). Another notable feature of 1D EP condensates arises from the compact nature of the phase, which, in contrast with the unbounded interface, allows for the generation of phase defects in space and time in the form of quantized $2\pi$ jumps often referred to as 
{spacetime vortices} \cite{Diehlspacetimevortex, Bloch&co2021, vercesi2023phase}.

In this work, we focus on a weakly-interacting 1D EP
Bose-Einstein condensate in the KPZ regime, where the phase is defect-free, and we investigate the blueshift of the energy spectrum. The dominant contributions to this energy shift, signature of the $U(1)$ symmetry breaking of the condensation transition, are the polariton-polariton and exciton reservoir-polariton mean-field interactions \cite{love2008blueshift, 2020ObersvationQDEPBEC}. 
Our aim is to determine the beyond-mean-field corrections to the blueshift induced by  fluctuations. The interest is two-fold: on the one hand, it provides a further characterization of  the difference between equilibrium and non-equilibrium condensates, and on the other hand, it deepens the understanding of the KPZ mapping. Let us elaborate on these two points. For an equilibrium quasi-BEC, the quantity analogous to the blueshift is the chemical potential, and its corrections associated to quantum and  thermal fluctuations  are well known \cite{LeeHuangYang57} (see also \cite{salasnich2018self}).
In non-equilibrium conditions, because the drive and loss are intrinsically stochastic  processes, the dominant fluctuations at weak interactions are of stochastic nature. We 
determine their effect on the blueshift, both through an analytical expression obtained within Bogoliubov approximation, and through numerical simulations of the full non-linear stochastic dynamics.

The KPZ mapping shows that within certain general assumptions, the dynamics of the condensate phase  follows the KPZ equation which describes the stochastic growth of a classical interface \cite{KPZ1986}. The KPZ height field $h(x,t)$ is known to behave at late times at a given space point as $h(t) \sim v_\infty t + (\Gamma t)^{1/3}\chi$, where $\Gamma$ is a non-universal parameter,  $v_\infty=\underset{t\to +\infty}{\lim}\langle \partial_t h\rangle$  is the asymptotic deterministic velocity of the interface \cite{Krugasymptoticvelocity}, and $\chi$ is a stochastic variable, whose distribution is non-Gaussian and is sensitive to the global geometry of the growth, or equivalently to its initial conditions. For a flat interface, the distribution of $\chi$ is a Tracy-Widom GOE (Gaussian orthogonal ensemble) distribution, while for a curved interface (droplet growth) it is a Tracy-Widom GUE (Gaussian unitary ensemble) distribution, and finally for the stationary growth it is the Baik-Rains distribution \cite{corwin2012KPZ}. This unveils a striking connection with random matrix theory where these distributions first emerged. The KPZ mapping  for the condensate implies that -- once the mean-field evolution is subtracted -- the unwrapped phase should also evolve as $\Tilde{\theta}(t) \sim -\Omega_\infty t + (\Gamma t)^{1/3}\chi$ where the minus sign is added to define $\Omega_\infty>0$ (see below). While much effort has been devoted to determining the statistics of the random variable $\chi$ in exciton-polariton systems  \cite{squizzato2018KPZsubclasses, deligiannis2021KPZsubclasses,Bloch&co2021}, the linear term  has received no attention yet. In particular, it was shown that the distinct distributions expected for a KPZ growth in the different geometries also arise for the phase fluctuations of exciton-polaritons, which is a definite evidence for the  validity of the mapping. In contrast, nothing is known on  $\Omega_\infty$, although this linear term should induce corrections, originating in the non-linear KPZ behavior, to the dominant mean-field contribution to the blueshift. This is the  issue addressed in this work. We show that the stochastic corrections to the blueshift are both due to density and phase fluctuations
 and hence go beyond  the effective KPZ dynamics, where only phase fluctuations are relevant.

The paper is organized as follows. The model for the stochastic dynamics of the EP condensate is presented in Sec.~\ref{sec:Intro_model}, and the link between the phase dynamics and the KPZ physics is detailed  in Sec.~\ref{sec:Intro_mapping}. We discuss in Sec.~\ref{sec:Intro_eqBEC} the connections between blueshift stochastic fluctuations and chemical potential quantum corrections. We determine in Sec.~\ref{sec:Bogo} the expression for the blueshift stochastic correction obtained within the Bogoliubov approximation. Our results from numerical simulations of the full dynamics are then presented in Sec.~\ref{sec:Num}, where we determine the dependence of this correction on the noise amplitude and on the interaction strength. Finally, Sec.\ref{sec:conclusions} offers some concluding remarks.

\section{Model and observables}
\subsection{\label{sec:Intro_model}
  Model for the dynamics of a 1D exciton-polariton condensate}

The dynamics of a weakly-interacting 1D exciton-polariton condensate subjected to 
incoherent pumping  can be modeled by a generalized Gross-Pitaevskii equation (gGPE) for the condensate wavefunction $\Phi$, coupled to a rate equation for the dynamics of the density of the excitonic reservoir $n_R$, as \cite{wouters2009stochasticGPE}
\begin{eqnarray}
i\hbar\partial_t\Phi & = & \Bigg[ \mathcal{F}^{-1}\left[\epsilon_{\hat{k}} - \dfrac{i\hbar}{2}\gamma_{\ell}(\hat{k}) \right] + \dfrac{i\hbar R}{2}n_R \nonumber\\
    & & + \hbar g|\Phi|^2 + 2\hbar g_Rn_R  \Bigg]\Phi + \hbar\xi \label{eq:gGPE} \\
\partial_tn_R & = & P - \left(\gamma_R + R|\Phi|^2 \right)n_R \label{eq:xreservoir}\,,
\end{eqnarray}
where $\mathcal{F}^{-1}$ denotes the inverse Fourier transform, with $\epsilon_{\hat{k}}=\hbar^2k^2/(2m)$ the  dispersion relation on the lower-polariton branch, and $\gamma_{\ell}(\hat{k})=\gamma_0+\gamma_2k^2$ the $k$-dependent polariton loss rate around the condensation momentum $k=0$, both within the quadratic approximation. The momentum dependence of the loss rate is typically observed in experimental realizations of polariton condensates \cite{Bloch&co2021}. $m$ denotes  the polariton mass, $g$ represents the polariton-polariton interaction strength and $g_R$  the exciton-polariton one. The excitonic reservoir is incoherently driven at pumping rate $P$, scatters with polaritons with amplitude $R$, and dissipates through other decay channels at rate $\gamma_R$. The complex noise $\xi$ is chosen of amplitude $\sigma$ with zero mean $\langle \xi \rangle=0$ and a covariance $\langle \xi(x,t)\xi^*(x',t') \rangle=2\sigma\delta(x-x')\delta(t-t')$.

\subsection{\label{sec:Intro_mapping}Mapping to the KPZ equation and asymptotic behavior of the phase}

It was realized that the dynamics of the phase of the driven-dissipative condensate maps within some conditions onto the KPZ equation \cite{altmanPRX2015KPZmapping,Ji2015}. 
In order to show this, one writes the condensate wavefunction $\Phi$  in density-phase representation $\Phi = \sqrt{n}e^{i\theta}$ and deduce from the gGPE~\eqref{eq:gGPE} the coupled  dynamical equations for the phase $\theta(x,t)$ and the density $n(x,t)$ fields as
\begin{align}
\partial_t n  =&  -\dfrac{\hbar}{m}\big(\partial_x n\partial_x\theta  + n \partial_x^2 \theta\big) \nonumber\\
 &+\gamma_2 \Big( \dfrac{1}{2}\partial^2_x n- \frac{\left( \partial_x n \right)^2}{4n} -n \left( \partial_x\theta \right)^2\Big)\nonumber\\
 &-\gamma_0 n + R n_R n -2 \sqrt{\sigma n} {\rm Im}\left(\xi e^{-i \theta}\right)    \label{eq:density_dynamics}\\
 \partial_t \theta  =&  \dfrac{\hbar}{2m}\left( \dfrac{1}{2}\dfrac{\partial^2_xn}{n} - \frac{1}{4}\left( \dfrac{\partial_x n}{n}\right)^2 - \left( \partial_x\theta \right)^2 \right) \nonumber \\
    & + \dfrac{\gamma_2}{2} \left(\dfrac{\partial_x n\partial_x \theta}{n} + \partial_x^2\theta\right) \nonumber\\
    &- g n - 2 g_R n_R +\sqrt{\dfrac{\sigma}{n}}{\rm Re}\left(\xi e^{-i \theta}\right) \, .
    \label{eq:phase_dynamics}
\end{align}
Then one expands the condensate and reservoir densities, as well as the condensate phase, in terms of small stochastic corrections around their mean-field noise-independent values as
\begin{eqnarray}
    n(x,t) & = & n_0 - \Tilde{n}(x,t)  \label{eq:expansion_n}\\
    n_R(x,t) & = & n_{R_0}+ \Tilde{n}_R(x,t) \label{eq:expansion_nR}\\
    \theta(x,t) & = & \theta_0(t)+ \Tilde{\theta}(x,t) \label{eq:expansion_theta},
\end{eqnarray}  
where the minus sign in \eqref{eq:expansion_n} is added to define $\langle \Tilde{n} \rangle>0$ in the following, and where $n_0=\gamma_R(P/P_{\text{th}}-1)/R$, $n_{R_0}=\gamma_0/R$, with the pump threshold for condensation $P_{\text{th}}=\gamma_R\gamma_0/R$. 
 The mean-field component of the phase evolves with time as $\theta_0(t)=-\Omega_0t$, where $\Omega_0=gn_0+2g_Rn_{R_0}>0$. This is a good approximation for the blueshift of the weakly-interacting EP condensate from which the interaction strength can be estimated \cite{love2008blueshift, 2020ObersvationQDEPBEC}.

Moreover,  one assumes that both density and reservoir fluctuations evolve adiabatically compared to the phase fluctuations $\Tilde{\theta}$, such that one can  neglect their temporal variations $\partial_t\Tilde{n}\approx0$, $\partial_t\Tilde{n}_R\approx0$. This yields for the reservoir density fluctuations $\Tilde{n}_R =- 2\frac{g_{i}}{R}\Tilde{n}$ with $g_{i}=(Rn_{R0})^2/2P$, 
and allows one to extract from~\eqref{eq:density_dynamics} the expression for the condensate density fluctuations $\Tilde{n}$ in terms of derivatives of $\Tilde{n}$ and of $\Tilde{\theta}$, which are then inserted  into Eq.~\eqref{eq:phase_dynamics} for the phase dynamics. 

By assuming that the density profile remains smooth, such that its spatial variations can be neglected, one finds that the phase is governed  by the following effective dynamical equation
\begin{eqnarray}
\partial_t\Tilde{\theta} = \nu\partial_x^2\Tilde{\theta} + \dfrac{\lambda}{2}( \partial_x\Tilde{\theta} )^2 + \sqrt{D}\eta \, ,
\label{eq:KPZmapping}
\end{eqnarray}
which is the KPZ equation \cite{KPZ1986}, where $\eta$ is a real Gaussian noise of zero mean $\langle \eta \rangle=0$ and delta correlations in space and time $\langle \eta(x,t)\eta(x',t') \rangle = 2\delta(x-x')\delta(t-t')$. Its parameters are given in terms  of the gGPE microscopic parameters by
\begin{align}
 \nu &= \frac{\gamma_2}{2} + \frac{\hbar g_{e}}{2mg_{i}} \label{eq:paramKPZ_nu} \\
 \lambda &= -\frac{\hbar}{m} + \gamma_2 \frac{g_{e}}{g_{i}}\label{eq:paramKPZ_lambda} \\
 D &= \frac{\sigma}{2n_0}\left[ 1+ \left(\frac{g_{e}}{g_{i}}\right)^2 \right], \label{eq:paramKPZ_D}
\end{align}
where the effective polariton-polariton interaction strength within the adiabatic approximation is denoted $g_{e}=g - 4\frac{g_Rg_{i}}{R}$.

 The KPZ dynamics of the phase determines the coherence properties of the EP condensate. These properties are encoded in the first-order correlation function of the condensate, defined as  
 \begin{equation}
 g^{(1)}(x,t)=\frac{\langle \Phi^*(x,t)\Phi(0,0) \rangle}{\sqrt{\langle n(x,t) \rangle\langle n(0,0) \rangle}}\,
 \end{equation}
 where the average $\langle\cdot\rangle$ is performed over realizations of the stochastic dynamics \eqref{eq:gGPE}.
 If one neglects the density-phase correlations, and assumes constant density-density correlations, then one obtains within a cumulant expansion  \cite{Bloch&co2021} that
 \begin{align}
 -2\log\left[ |g^{(1)}(x,t)| \right] &\simeq \Big\langle \big(\Tilde{\theta}(x,t) - \Tilde{\theta}(0,0)\big)^2 \Big\rangle\nonumber\\
 & \equiv  C_{\theta\theta}(x,t)
 \label{eq:cumulant}
 \end{align}
where $C_{\theta \theta}$ is the connected two-point correlation function of the phase. We adopt in the following the short-hand notation $\mathcal{G}^{(1)}(x,t)\equiv -2\log\left[ |g^{(1)}(x,t)| \right]$.
For the KPZ dynamics of the height field $h(x,t)$, the connected correlation function 
 $C^{\rm KPZ}_{hh}(x,t)$ is known to take a scaling form given by 
\begin{eqnarray}
C^{\rm KPZ}_{hh}(x,t)= C_0 t^{2/3} g_{\rm KPZ}\left(y_0\dfrac{x}{t^{2/3}}\right),
\label{eq:scaling_form_G1_KPZ}
\end{eqnarray}
from which we read the 1D KPZ universal growth and roughness critical exponents $\beta_{\rm KPZ}=1/3$ and $\alpha_{\rm KPZ}=1/2$ respectively. $C_0$, $y_0$ are non-universal constants, and $g_{\rm KPZ}$ is the KPZ universal scaling function, which was calculated exactly in \cite{prahoferspohn2004}. It has the  following asymptotic behaviors: $g_{\rm KPZ}(y) \to g_0$, with  $g_0$ finite, when $y\to 0$ and  $g_{\rm KPZ}(y) \sim |y|$ when $y\to \infty$. Hence, the correlation function behaves as power-laws  both at  coinciding space-points $C^{\rm KPZ}_{hh}(0,t)\sim  t^{2\beta_{\rm KPZ} } \equiv t^{2/3}$ and equal times $C^{\rm KPZ}_{hh}(x,0)\sim |x|^{2\alpha_{\rm KPZ}}\equiv |x|$. 

The linearized version of the KPZ equation \eqref{eq:KPZmapping} (setting $\lambda=0$) is the Edwards-Wilkinson (EW) equation. Its corresponding height field connected correlation function $C_{hh}^{\rm EW}(x,t)$ takes a similar scaling form
\begin{eqnarray}
C^{\rm EW}_{hh}(x,t)= C_1 t^{1/2} g_{\rm EW}\left(y_1\dfrac{x}{t^{1/2}}\right),
\label{eq:scaling_form_G1_EW}
\end{eqnarray}
where $C_1,y_1$ are non-universal constants. The universal exponents $\beta_{\rm EW}=1/4$, $\alpha_{\rm EW}=1/2$, and the universal scaling function $g_{\rm EW}$ define the EW universality class \cite{edwards1982surface, takeuchi2018appetizer}.

\begin{figure}[h!]
\includegraphics[width=7cm]{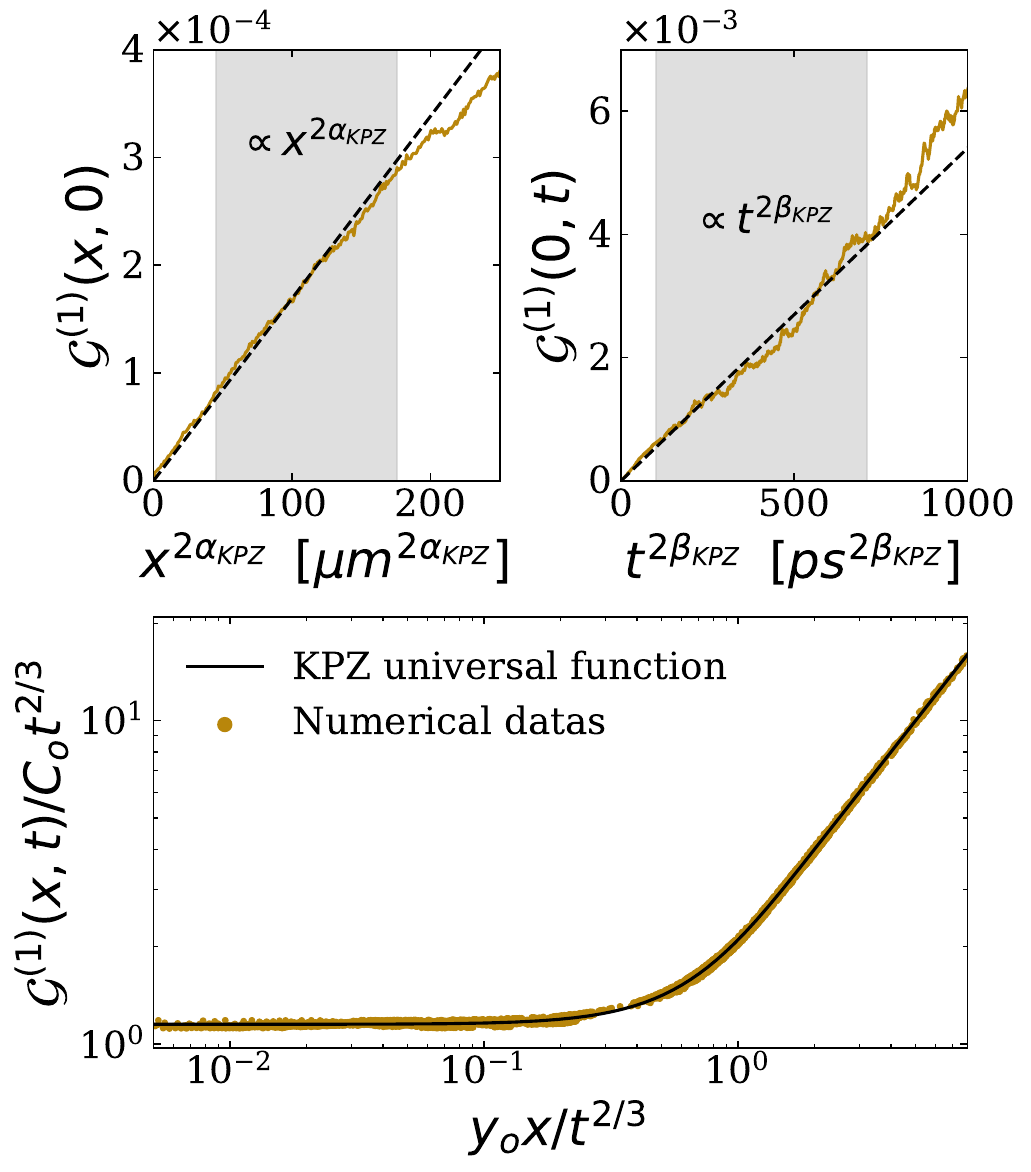}
\caption{\label{fig:KPZscaling} KPZ scaling regime: (upper left panel) spatial scaling of $\mathcal{G}^{(1)}(x,0)$ as a function of $x$, indicated by the grey shade in between $\left[60 \mu m, 220 \mu m \right]$ ; (upper right panel) temporal scaling of $\mathcal{G}^{(1)}(0,t)$ as a function of $t^{2/3}$, indicated by the grey shade in between $\left[10^{3} ps, 2\times10^4 ps \right]$; (lower panel) collpase of $\mathcal{G}^{(1)}(x,t)$ onto the KPZ universal scaling function $g_{\rm KPZ}$. The parameters used on this figure are $\gamma_R=4.5\gamma_0$, and $\sigma=10^{-2}\gamma_0/2$. The system size is set to $2000 \mu m$. Other parameters are specified in Section \ref{sec:Num}.}
\end{figure}

\bigbreak 
 It follows  from \eqref{eq:cumulant} that if the phase of the condensate
 obeys a KPZ dynamics at least on certain time and length scales, with an effective non-linearity $\lambda\neq0$, then the coherence of the condensate endows the  scaling form 
\begin{equation}
\mathcal{G}^{(1)}(x,t) \simeq C^{\rm KPZ}_{hh}(x,t)\sim \left\{\begin{array}{l l}
                                            t^{2/3} \qquad& x=0\\
                                            |x| & t=0
                                        \end{array}\right. \,,
                                           \label{eq:KPZscaling}
\end{equation}
in the space-time window where the KPZ dynamics is relevant. 
This behavior has been recently reported in experiments  \cite{Bloch&co2021}, which confirms the emergence of KPZ universality in exciton-polariton condensates. We also find a KPZ regime in our numerical simulations, reported in Sec.~\ref{sec:Num} (and Fig.~\ref{fig:KPZscaling}).

As mentioned in the introduction, another hallmark of KPZ universality is the probability distribution of the reduced height fluctuations $\chi$ defined from the  long time behavior at a given space point as $h(t) \sim v_\infty t + (\Gamma t)^{1/3} \chi$.  In the absence of phase defects, $\Tilde{\theta}$ can be safely unwrapped into a continuous variable, which is expected in the KPZ regime to follow the asymptotic form 
\begin{equation}
 \Tilde{\theta}(x_0,t) \underset{t\to +\infty}{\sim}-\Omega_{\infty}t + (\Gamma t)^{1/3}\chi\, .
 \label{eq:omegainfanz}
\end{equation}
 The first term  describes the non-universal linear growth of the phase fluctuations through the asymptotic frequency defined as $\Omega_{\infty}=-\underset{t\to +\infty}{\lim}\langle\partial_t\Tilde{\theta}\rangle$. In the second term, the power-law $t^{1/3}$ gives rise to the  KPZ universal scaling law for $C_{\theta\theta}$ in time, and the distribution of the fluctuations $\chi$ was shown to exhibit the  Tracy-Widom and Baik-Rains forms expected for the KPZ universality class, where the curved interface geometry in particular can be emulated for the phase through a confinement potential \cite{deligiannis2021KPZsubclasses}.

 In this work, we focus on the correction  $\Omega_{\infty}$,
 due to stochastic fluctuations, to the mean-field blueshift $\Omega_0$. 
 We investigate its dependence on the system parameters using Bogoliubov theory and numerical simulations of Eqs.~\eqref{eq:gGPE} and \eqref{eq:xreservoir}. Let us first elaborate on the link between the blueshift in non-equilibrium condensates and  the analogous quantity defined at equilibrium.

\subsection{\label{sec:Intro_eqBEC}Connection with equilibrium quasi-condensates}
For equilibrium weakly-interacting Bose-Einstein condensates, in homogeneous conditions and with contact interactions $v(x-x')=\hbar g \delta (x-x')$, the mean-field time evolution of the wave-function  $\Phi(t)=\sqrt{n}e^{i\theta(t)}$ is only contained in the phase and is linear in time $\theta(t)=-\frac{\mu_0}{\hbar} t$, where the blueshift $\mu_0/\hbar =gn_0$ is the condensate chemical potential.
The beyond mean-field quantum
corrections to the blueshift
at weak interactions 
have been calculated  for a 3D
Bose gas by Lee, Huang and Yang \cite{LeeHuangYang57} within the Bogoliubov approximation in terms of the scattering length of the fluid. Ref.~\cite{salasnich2018self} provides an alternative derivation of such corrections based on an equation of motion approach, yielding the following expression for  the chemical potential $\mu/\hbar = gn_0 + 2 g\langle \Tilde{n} \rangle_Q +  g\langle \tilde{m} \rangle_Q$, where $\langle  \cdot \rangle_Q$ denotes the
quantum average over the ground state wavefunction,  $\langle \Tilde{n} \rangle_Q$ is the non-condensed particle density 
corresponding to  the quantum depletion at zero temperature, and $\langle \Tilde{m} \rangle_Q$ is the anomalous density.
The above calculation has also been extended to
low dimensional quasi-condensates \cite{lieb1963exact,schick1971,popov1972,popov-book1983}.

For the non-equilibrium 1D EP condensate, one also finds that, at the mean-field level, the phase linearly decreases with time with the blueshift $\Omega_0=gn_0+2g_Rn_{R0}$, which is analogous to the mean-field expression for the chemical potential $\mu_0$.
The total blueshift $\Omega = \Omega_0+\Omega_{\infty}$ can thus be interpreted as the equivalent of the chemical potential for the equilibrium condensate. However, since the EP condensate is out-of-equilibrium, $\Omega$  cannot be defined by the usual  thermodynamics relation $\partial E/\partial N\large|_{T,V}$, but instead has to be calculated from the dynamics $\Omega =-\langle\partial_t\theta\rangle$ similarly to Ref.~\cite{salasnich2018self}. The term $\Omega_{\infty}$ is hence  analogous to the quantum Lee-Huang-Yang corrections at equilibrium, except that here it originates from  stochastic fluctuations.

\section{\label{sec:Bogo}Bogoliubov approximation in the adiabatic limit of the excitonic reservoir}

In this section, we derive an expression for the blueshift correction $\Omega_{\infty}$ using the linearized Bogoliubov approximation.

\subsection{\label{sec:Bogo_adiab}Linearized dynamics and equal time correlation functions}

Since the time evolution of the reservoir dynamics is typically much slower  
than the time scales of the condensate dynamics,
we neglect it in the following. Within this adiabatic approximation, we express the reservoir density in terms of the condensate density and hence close  the set of coupled equations Eqs.~\eqref{eq:density_dynamics} and \eqref{eq:phase_dynamics}.
Linearizing these equations gives the coupled dynamics for  the fluctuations  of density $\delta n$ and phase $\delta \theta$ to first order in the noise. 
This reads in Fourier space as
\begin{eqnarray}
    \begin{pmatrix}
     \delta n/2n_0\\
    \delta \theta
    \end{pmatrix}(k,\omega) = G(k,\omega)\begin{pmatrix}
\Tilde{\xi}_1 \\
\Tilde{\xi}_2
\end{pmatrix}
\label{eq:GPElinearized}
\end{eqnarray}
where $G(k,\omega) = \left( i\omega - \mathcal{L}_{rot} \right)^{-1}$ is the Green function, $\mathcal{L}_{rot}$ is the adiabatic Bogoliubov matrix
\begin{equation}
    \mathcal{L}_{rot} = \begin{pmatrix}
        \Gamma_{\hat{k}}-g_{i}n_0   &  \epsilon_{\hat{k}}  \\
        -\left( \epsilon_{\hat{k}} + 2g_{e}n_0 \right) & \Gamma_{\hat{k}}+g_{i}n_0 
    \end{pmatrix}
    \label{eq:Lrot_general}
\end{equation}
with $\Gamma_{\hat{k}} = \frac{\Delta\gamma(\hat{k})}{2} - g_{i}n_0$, $\Delta\gamma(\hat{k}) = Rn_{R0}-\gamma_{\ell}(\hat{k})$ and where the noise satisfies $\langle \Tilde{\xi}_i \rangle = 0$, $\langle \Tilde{\xi}_1\Tilde{\xi}_2 \rangle = 0 $, $\langle \Tilde{\xi}_i(k,\omega)\Tilde{\xi}_i(k',\omega') \rangle = 4\pi^2\dfrac{\sigma}{n_0} \delta(k+k')\delta(\omega+\omega')$.

Within our model for $\gamma_{\ell}(\hat{k})$ and $\epsilon_{\hat{k}}$ (see Sec.~\ref{sec:Intro_model}), the Bogoliubov matrix of Eq.~\eqref{eq:Lrot_general} simplifies to 
\begin{eqnarray}
    \mathcal{L}_{rot}=
    \begin{pmatrix}
        - 2g_{i}n_0 - \dfrac{\gamma_2}{2}k^2 & \dfrac{\hbar k^2}{2m} \\
        -\dfrac{\hbar k^2}{2m} - g_{e}n_0 & - \dfrac{\gamma_2}{2}k^2
    \end{pmatrix}.
\end{eqnarray}
From Eq.~\eqref{eq:GPElinearized}, we find the equal time correlation functions, expressed  in terms of both the gGPE microscopic parameters and the effective viscosity $\nu$ and nonlinearity $\lambda$ of the KPZ mapping (Eqs.~\eqref{eq:paramKPZ_nu} and \eqref{eq:paramKPZ_lambda} respectively) 
\begin{eqnarray}
    \langle \delta n(k) \delta n(-k) \rangle & = & \dfrac{2\sigma n_0}{\gamma_2} \dfrac{2k^2 + \kappa_h^2}{\left( k^2 + \kappa_h^2 \right)\left( k^2 + \kappa_d^2 \right)} \label{eq:nn_k}\\
    \langle \delta n(k) \delta \theta(-k) \rangle & = & -\dfrac{\sigma\kappa_h^2n_0g_{i}\lambda}{\nu\gamma_2^2\kappa_d^2\left( k^2 + \kappa_h^2 \right)\left( k^2 + \kappa_d^2 \right)} \label{eq:ntheta_k} \\
    \langle \delta \theta (k) \delta \theta(-k) \rangle & = &  \dfrac{\langle \delta n(k) \delta n(-k) \rangle}{4n_0^2} + \dfrac{\sigma \kappa_h^2}{\nu n_0\gamma_2^2\kappa_d^2} \nonumber \\
    &&  \times \dfrac{2\nu g_{i}n_0k^2 + 2n_0^2\left(g_{i}^2 + g_{e}^2\right)}{k^2\left( k^2 + \kappa_h^2 \right)\left( k^2 + \kappa_d^2 \right)} \label{eq:thetatheta_k} ,
\end{eqnarray}
where we introduced two momentum scales $\kappa_h$ and $\kappa_d$ representing the non-equilibrium healing length and  dissipation scale, respectively, defined by  
\begin{equation}
\kappa_h=\sqrt{\dfrac{8\nu g_{i}n_0}{\hbar^2/m^2 + \gamma_2^2}}\, ,\quad \kappa_d=\sqrt{\dfrac{2g_{i}n_0}{\gamma_2}}\, ,
\end{equation}
with typically $\kappa_h>\kappa_d$. Note that the choice of $\gamma_{\ell}(\hat{k})$ we made regulates the large-momentum behavior of Eqs.~\eqref{eq:nn_k} and \eqref{eq:thetatheta_k} through the effective loss rate $\Gamma_{\hat{k}}$, similarly to the frequency dependent amplification introduced in Ref.~\cite{chiocchetta2013}. Generic expressions of the equal time correlation functions for unspecified $\gamma_{\ell}(\hat{k})$ and  $\epsilon_{\hat{k}}$ are given in Appendix~\ref{sec:ap_bogo}. The effect of alternative choices for the polariton loss rate $\gamma_{\ell}(\hat{k})$ is discussed in Appendix~\ref{ap:k-dependent_dissipation}.

In addition to $C_{\theta\theta}(x)\equiv C_{\theta\theta}(x,0)$ defined in Eq.~\eqref{eq:cumulant}, we introduce the density-density and density-phase equal time correlation functions  $C_{nn}(x)=\langle  \delta n(x)\delta n (0) \rangle - \langle  \delta n(0)^2 \rangle$ and  $C_{n\theta}(x)=\langle  \delta n(x)\delta \theta (0) \rangle$ respectively. Within Bogoliubov approximation, we find from Eqs.~\eqref{eq:nn_k}-\eqref{eq:thetatheta_k} that
\begin{eqnarray}
    C_{nn}(x) & = &  \dfrac{\sigma n_0}{\gamma_2} \Bigg[ \dfrac{\kappa_h }{\kappa_h^2-\kappa_d^2}\left(e^{-\kappa_h|x|} - 1 \right) \nonumber \\
    && + \left( 2 - \dfrac{\kappa_h^2}{\kappa_h^2 - \kappa_d^2} \right)\dfrac{e^{-\kappa_d|x|}-1}{\kappa_d} \Bigg] \label{eq:Cnn} \\
    C_{n\theta}(x) & = & \sigma\lambda \kappa_h^2 \dfrac{e^{-\kappa_h|x|}/\kappa_h - e^{-\kappa_d|x|}/\kappa_d}{4\gamma_2 \nu \left(\kappa_h^2-\kappa_d^2\right)} \label{eq:Cnheta} \\
    C_{\theta\theta}(x) & = &  -\dfrac{C_{nn}}{2n_0^2} + \dfrac{e^{-{\kappa_d|x|}}-1}{\kappa_d/\mathcal{K}^{(d,d)}}  \nonumber \\
    &&  - \dfrac{e^{-{\kappa_h|x|}}-1}{\kappa_h/\mathcal{K}^{(h,h)}} + \dfrac{D}{\nu \kappa_d^2}|x| \label{eq:Cthetatheta},
\end{eqnarray}
where $\nu$, $\lambda$, $D$ are the KPZ parameters introduced in Eqs.~\eqref{eq:paramKPZ_nu}-\eqref{eq:paramKPZ_D}, and with the momentum matrix elements
\begin{eqnarray}
    \mathcal{K}^{(i,j)} & = & \dfrac{\kappa_h^2}{\kappa_h^2 - \kappa_d^2}\left[ \dfrac{D\kappa_d^2}{\nu \kappa_i \kappa_j} - \dfrac{\sigma}{\gamma_2 n_0} \right] \label{eq:Kfct_k}
\end{eqnarray}

Let us comment on the phase-phase correlation function Eq.~\eqref{eq:Cthetatheta}.
We observe that at long distances, it exhibits the scaling   $C_{\theta\theta}\propto |x|$, which reveals the deep connection between the condensate phase dynamics and stochastically growing  interfaces (see Appendix~\ref{ap:higher_dim}), leading to $|g^{(1)}(x,0)|\sim e^{-|x|}$. This scaling indeed corresponds to the expected one for the equal-time EW 
correlation \eqref{eq:scaling_form_G1_EW} with $\alpha=1/2$. The same equilibrium spatial scaling is obtained for the KPZ dynamics, merely reflecting that the stationary KPZ interface is a Brownian in space. { This spatial scaling coincides with the one of 1D equilibrium quasi-condensates at finite temperature \cite{giamarchi-book}, as mentioned in the introduction.}
It should also be noted that $C_{\theta\theta}$ explicitly depends on $C_{nn}$. While being irrelevant at large distances as mentioned above, this plays an important role when $x\to0$ and will be crucial in the following.

\subsection{\label{sec:Bogo_corr_sd_Omegainfty}Blueshift corrections}

We obtain the blueshift $\Omega_\infty =  -\langle \partial_t\Tilde{\theta} \rangle$
by taking the average of Eq.~\eqref{eq:phase_dynamics} over the stochastic fluctuations. This average involves on the r.h.s spatial derivatives of two-point correlation functions of the phase and density, which we approximate by their expressions within Bogoliubov theory \eqref{eq:Cnn}-\eqref{eq:Cthetatheta}. 
For instance, we have $\langle (\partial_x\Tilde{\theta})^2 \rangle \approx \langle \left(\partial_x\delta\theta\right)^2 \rangle = \frac{1}{2}\underset{x\to 0}{\lim}\partial_x^2C_{\theta\theta}(x)$.
Hence, the blueshift correction reads
\begin{eqnarray}
    \Omega_{\infty }& \underset{x\to 0}{=} & \dfrac{\hbar}{4m}\left[ \dfrac{\partial_x^2C_{nn}}{2n_0^2} + \partial_x^2C_{\theta\theta}\right] \nonumber \\
   &&  + \gamma_2 \dfrac{\partial_x^2C_{n\theta}}{2n_0} - g_{e}\langle \Tilde{n} \rangle_S
    \label{eq:Omega_infty_CAB}
\end{eqnarray}
where $\langle \Tilde{n} \rangle_{S} \equiv \langle \Tilde{n} \rangle$ is the non-condensate density arising from stochastic fluctuations. 
Similarly to $\Omega_{\infty}$, it is obtained 
from the dynamics of the density fluctuations $\Tilde{n}$ by inserting the expansions \eqref{eq:expansion_n}-\eqref{eq:expansion_theta} into Eq.~\eqref{eq:density_dynamics}, performing the adiabatic approximation of the reservoir dynamics. Assuming stationarity $\partial_t \Tilde{n} =0$  and taking the average, we can extract $\langle \Tilde{n} \rangle_S$ in terms of spatial derivatives of two-point correlation functions. Following the same steps as above and approximating these correlation functions by Eq.~\eqref{eq:Cnn}-\eqref{eq:Cthetatheta} gives 
\begin{eqnarray}
    \langle \Tilde{n} \rangle_{S} & \underset{x\to 0}{=} &  \dfrac{\gamma_2}{4g_{i}} \left[ \dfrac{\partial_x^2 C_{nn}}{2n_0^2} + \partial_x^2C_{\theta\theta} \right]  - \dfrac{\hbar\partial_x^2C_{n\theta}}{2g_{i}mn_0}.
    \label{eq:stochastic_depletion_CAB}
\end{eqnarray}
Replacing $\langle \tilde{n} \rangle_S$ in Eq.~\eqref{eq:Omega_infty_CAB}, one finds an expression of $\Omega_{\infty}$ which can be conveniently expressed in terms of the KPZ parameters $\lambda$ and $\nu$ as
\begin{eqnarray}
    \Omega_{\infty }& \underset{x\to 0}{=} & -\dfrac{\lambda}{4}\left[ \dfrac{\partial_x^2C_{nn}}{2n_0^2} + \partial_x^2C_{\theta\theta}\right] + \nu \dfrac{\partial_x^2C_{n\theta}}{n_0}. \label{eq:Omega_infty_CAB_lambda}
\end{eqnarray}

It is instructive to compare this expression, obtained from the coupled density-phase dynamics Eq.~\eqref{eq:GPElinearized}, with the one ensuing from the KPZ equation of phase fluctuations Eq.~\eqref{eq:KPZmapping} alone. For the latter, one can follow the same procedure,  {\it i.e.} one first linearizes the dynamics Eq.~\eqref{eq:GPElinearized}, which yields the EW equation, then computes the associated correlation function $C^{\rm EW}_{\theta\theta}(x)$, and finally defines  $\Omega_{\infty}$ as $\Omega_{\infty}=-\underset{t\to +\infty}{\lim}\langle \partial_t\Tilde{\theta}\rangle$ from the full KPZ dynamics \eqref{eq:KPZmapping}. One then finds   $\Omega_{\infty} \underset{x\to0}{=} -\frac{\lambda}{4}\partial_x^2C^{\rm EW}_{\theta\theta}(x)$. 
While this term corresponds to the second one in \eqref{eq:Omega_infty_CAB_lambda}, additional terms enter the latter expression, which thus does not depend on the phase only.
 This means that, even though the spatio-temporal large scale properties of the quasi-condensate are satisfactorily described by the effective KPZ dynamics \eqref{eq:KPZmapping} (see Fig.~\ref{fig:KPZscaling} and Refs.~\cite{squizzato2018KPZsubclasses, deligiannis2021KPZsubclasses, Bloch&co2021}), the correlations of the density, and the coupling between density and phase bring additional contributions to $\Omega_{\infty}$.
 
 Moreover, let us emphasize that the expression for $\Omega_{\infty}$ is sensitive to the microscopic details of the condensate dynamics
 stemming from the short-distance properties of both density and phase correlation functions \eqref{eq:Cnn}-\eqref{eq:Cthetatheta}. In fact,  both $C_{nn}$ and $C_{\theta\theta}$ contributions to \eqref{eq:Omega_infty_CAB_lambda} are needed in order to make $\Omega_{\infty}$ well defined in the UV regime. Indeed, taking the second spatial derivative of  $C_{\theta\theta}(x)$, one finds that it diverges  for $x\to0$. While this is also the case for $C^{\rm EW,KPZ}_{hh}(x)$, the particularity here is that the diverging part comes from the first term in  the r.h.s of \eqref{eq:Cthetatheta}, i.e
\begin{eqnarray}
    \partial_x^2C_{\theta\theta} & \underset{x\to0}{=} & \left.-\dfrac{\partial_x^2C_{nn}}{2n_0^2}\right\rvert_{x=0} + \mathcal{K}^{(h,d)}\left( \kappa_h - \kappa_d\right)
    \label{eq:derivative_Cthetatheta_bogo}
\end{eqnarray}
where the density contribution reads
\begin{eqnarray}
    -\dfrac{\partial_x^2C_{nn}}{2n_0^2} & \underset{x\to0}{=} & \dfrac{\sigma}{2n_0\gamma_2}\left[ 4\dfrac{\Lambda}{\pi} - 2\kappa_d  \dfrac{\kappa_h^2}{\kappa_h+\kappa_d} \right]
    \label{eq:derivative_Cnn_bogo}
\end{eqnarray}
with $\Lambda$ an arbitrary UV cut-off.
 A similar divergence is also predicted for 1D quasi-BECs at equilibrium, where the standard regularization is to consider that Bogoliubov theory is applied on a lattice of spacing $\ell = \pi/\Lambda$. As formulated in Ref.~\cite{Castin2003extension}, the theory is valid as long as the parameter $\langle (\partial_x\delta\theta)^2 \rangle \propto 1/n_0\ell^3$ is small in 1D.  Adapted to the non-equilibrium dynamics, the same argument implies  that $\langle \left(\partial_x\delta\theta\right)^2\rangle \underset{x\to0}{=} \frac{1}{2}\partial_x^2C_{\theta\theta}(x)\propto \frac{\sigma}{n_0\gamma_2\ell}$ must remain a small parameter. As a complement, we show in Appendix~\ref{ap:higher_dim} that the criterion for the applicability of Bogoliubov theory on the driven-dissipative BEC can, as for its equilibrium counterpart, also be generalized to higher dimensions. In fact, we  show in Sec~\ref{sec:short_scales} that the dependence of the gradients of correlation functions on the short length-scale $\ell$ is not  an artefact of the linearization of \eqref{eq:density_dynamics} and \eqref{eq:phase_dynamics}, but also emerges at the level of the gGPE from the full non-linear dynamics of Eqs.~\eqref{eq:gGPE} and \eqref{eq:xreservoir}.

In the one-dimensional case for closed systems, because of the Mermin-Wagner-Hohenberg theorem, $\langle \tilde n \rangle_Q$ and $\langle \tilde m \rangle_Q$ both diverge in the IR. The two diverging terms are of opposite signs and cancel out, leading to a finite chemical potential correction \cite{Castin2003extension}. 
The same mechanism operates here,    the diverging contributions in the density and phase correlations precisely compensate in \eqref{eq:Omega_infty_CAB_lambda}, such that the  blueshift correction is finite, and reads, within the Bogoliubov approximation
\begin{eqnarray}
    \Omega_{\infty} & = & -\dfrac{\lambda D}{4\nu\left( \kappa_h^{-1} + \kappa_d^{-1} \right)}\label{eq:Omega_infty}
\end{eqnarray}
Let us  comment on this result. The correction $\Omega_{\infty}$ is found to be proportional to the ratio of KPZ effective parameters $\frac{\lambda D}{\nu}$ and in particular to the noise strength $\sigma$ through $D$. Additionally, as already emphasized, $\Omega_{\infty}$ depends on the microscopic details of the condensate through both the non-equilibrium healing length and dissipation length. 
It is interesting to remark that also  the linewidth of extended lasers depends on two-point correlation functions \cite{Exter_phase_diffusion, wenzel2021linewidth_theory_revisited, amelio2022Bogo_linewidth, amelio2023KPZlinewidth}, where $D$ is the Schawlow-Townes linewidth enhanced by the Henry factor $g_e/g_i$ \cite{ST1958linewidth_MASERS,henry1982theory}.

Let us discuss the sign of this blueshift correction. For the quasi-condensate to be stable, the mass $m$ and the effective coupling $g_{e}$ must have the same sign \cite{baboux2018}. This implies that $\nu>0$, while $\lambda$ can be either positive or negative, implying that the sign of $\Omega_{\infty}$ is fully controlled by the KPZ nonlinearity $\lambda$. This is consistent with the interpretation of $\Omega_{\infty}$ as the average growth velocity of the growing interface $\Tilde{\theta}$. In typical experimental realizations of stable exciton-polariton condensates, $m<0$ and $\lambda<0$ \cite{baboux2018, Bloch&co2021} so that $\Omega_{\infty}$ is positive.

Let us now comment on the correction $\langle \Tilde{n}\rangle_S$ to the condensate density. Replacing \eqref{eq:Cnn}-\eqref{eq:Cthetatheta} in \eqref{eq:stochastic_depletion_CAB}, we obtain
\begin{eqnarray}
    \langle \Tilde{n}\rangle_S & = & \dfrac{Dn_0}{2\nu\kappa_d^2\left( \kappa_h^{-1}+\kappa_d^{-1} \right)} + \dfrac{\sigma}{\gamma_2\left( \kappa_h + \kappa_d \right)},
    \label{eq:stochastic_depletion}
\end{eqnarray}
from which we read $\langle \Tilde{n}\rangle_S > 0$ for any parameters. 
The situation is analogous to the Bogoliubov occupation of three-dimensional BECs $\langle \Tilde{n}\rangle_Q$, which can only take positive values thus depleting the condensate, and reflecting the fact that the condensate chemical potential is slightly higher than its mean field value. There are however two main differences with the equilibrium case. First, $\langle \Tilde{n}\rangle_S$ is well defined in 1D while $\langle \Tilde{n}\rangle_Q$ diverges because of large phase fluctuations. Second, $\langle \Tilde{n}\rangle_S$ is not proportional to $\lambda$ and cannot change sign contrary to $\Omega_{\infty}$. Recalling that $\Omega_{\infty}$ represents an analogous chemical potential correction for the driven-dissipative condensate, this implies that, at variance with  equilibrium BECs, one could in principle reach a parameter regime for which the stochastic depletion $\langle \Tilde{n}\rangle_S$ is accompanied by a decrease of the total chemical potential. This situation cannot be realized for systems at equilibrium, where these two quantities are related by a state equation.

\bigbreak
In order to test the predictions \eqref{eq:Omega_infty}-\eqref{eq:stochastic_depletion} ensuing from the Bogoliubov linearized dynamics, we now perform numerical simulations of the full non-linear dynamics  Eqs.~\eqref{eq:gGPE} and \eqref{eq:xreservoir} and extract the numerical estimate for $\Omega_{\infty}$.  We study in particular its dependence on the KPZ noise strength $D$ through the gGPE noise $\sigma$, and we also vary the coupling strength $g$ to make the non-linearity $\lambda$ of the effective phase dynamics Eq.~\eqref{eq:KPZmapping} gradually decrease to zero.

\section{\label{sec:Num}Numerical simulations}

Numerical resolutions of Eqs.~\eqref{eq:gGPE} and \eqref{eq:xreservoir} are performed using a combination of split-step method and first-order Euler scheme \cite{agrawal2001}. The parameters are chosen to reproduce the  typical experimental conditions in GaAs microcavities \cite{Bloch&co2021}: $\hbar\gamma_0=48.5\mu eV$, $\hbar\gamma_2=1.6\times10^4\mu eV.\mu m^{2}$, $\gamma_R=0.45\gamma_0$, $R=8.8\times10^{-4}\mu m.ps^{-1}$, $P=1.1P_{\text{th}}$, $\hbar g_R=6\times10^{-4}\mu m.meV$. The system size is $500\mu m$ with lattice parameter $dx = 3.9\mu m$. The polariton mass $m=-3.3\times10^{-6}m_e$ is negative to obtain a stable condensate despite the attractive effective interaction between polaritons \cite{baboux2018}. The nominal value of the noise strength $\sigma$ is $\sigma_N=\frac{1}{2}\gamma_0$ and is controlled by the reservoir \cite{carusottoQFL, wouters2009stochasticGPE}. In order to test the validity of the expression  \eqref{eq:Omega_infty}, we consider the noise amplitude as a free parameter and we allow us to vary it independently of $\gamma_0$. The polariton-polariton interaction strength $g$ is set to zero in Sec.~\ref{sec:Num_fctsigma} and increased in Sec.~\ref{sec:Num_lambda}.

\subsection{Large-scale behavior: KPZ scaling regime}

For the chosen parameters, a KPZ regime was reported  both in experiments and numerical simulations \cite{Bloch&co2021}. In order to check this, we first compute  the first-order correlation function $g^{(1)}(x,t)$ of the condensate. The results for $\mathcal{G}^{(1)}(x,t)$ are shown in Fig.~\ref{fig:KPZscaling} for a system of length $2000\mu m$ and a smaller exciton lifetime $\gamma_R=4.5\gamma_0$, to enlarge the scaling windows for the sake of illustration.
We observe the expected scaling behavior \eqref{eq:KPZscaling} of both the equal-time and equal-space correlations with the KPZ exponents. Moreover, the whole set of space-time data points of  $\mathcal{G}^{(1)}(x,t)$ lying in the KPZ window perfectly collapses onto the exact KPZ universal scaling function. This confirms that the phase follows a KPZ dynamics in this space-time window.

\subsection{\label{sec:short_scales}Short-scale behavior: dependence on lattice spacing}

\begin{figure}[h!]
\includegraphics[width=8cm]{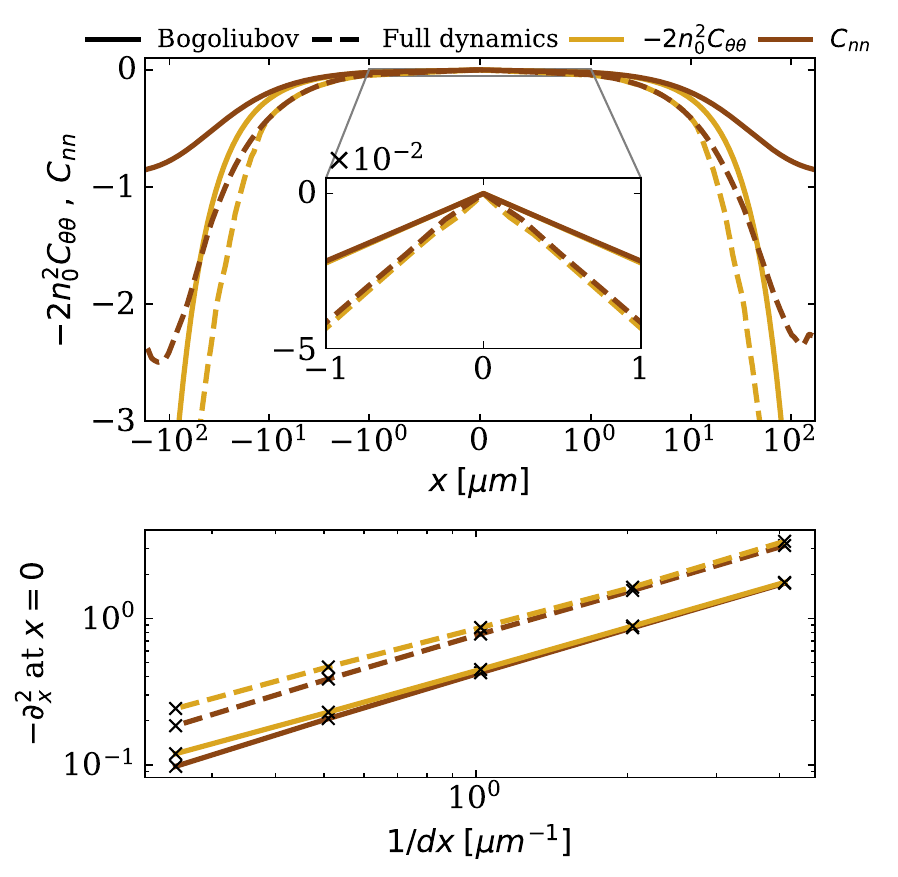}
\caption{\label{fig:short_scale_rho_theta} (upper panel) Spatial profile of $C_{nn}(x)$ (brown) and $-2n_0^2C_{\theta\theta}(x)$ (orange) for $dx=\ell=0.24 \mu m$ from Bogoliubov theory (solid line) and numerical resolution of the full dynamics (dashed line). (lower panel) Second derivatives evaluated at $x=0$ as a function of the inverse lattice spacing $1/dx$, same lines and color code  as in the upper panel.}
\end{figure}

We established in Sec.~\ref{sec:Bogo_corr_sd_Omegainfty} that, within the Bogoliubov approximation, the second derivative of the two-point correlation functions of the phase and density of the condensate both diverge at short distance, which requires the introduction of a UV cut-off $\Lambda \sim 1/\ell$. We here study this aspect,
computing numerically $C_{nn}$ and $C_{\theta\theta}$ from the full dynamics Eqs.~\eqref{eq:gGPE} and \eqref{eq:xreservoir} and 
for different values of the lattice spacing $dx$. The results obtained for $dx\approx0.24\mu m$ are shown on Fig.~\ref{fig:short_scale_rho_theta} (upper panel) and are found to be accurately described by Eqs.~\eqref{eq:Cnn} and \eqref{eq:Cnheta} from Bogoliubov approximation.

Second derivatives evaluated at coincident points $x=0$ of both $C_{nn}$ and $C_{\theta\theta}$ are shown for different lattice spacings on Fig.~\ref{fig:short_scale_rho_theta} (lower panel) together with the predictions of Eqs.~\eqref{eq:derivative_Cnn_bogo} and \eqref{eq:derivative_Cthetatheta_bogo}, where we set $\ell=dx$. We show that at $x=0$, $2n_0^2\partial_x^2C_{\theta\theta}$ closely follows $-\partial_x^2C_{nn}$ and evolves as $1/dx$, confirming the relations derived within Bogoliubov approximation. In fact, results obtained from the nonlinear dynamics merely differ from Eqs~\eqref{eq:derivative_Cnn_bogo}-\eqref{eq:derivative_Cthetatheta_bogo} by a prefactor.
Hence, this divergence is not a mere artefact of the Bogoliubov approximation, but is intrinsic to the non-equilibrium gGPE.
The divergences also compensate in the full dynamics such that the overall blueshift correction $\Omega_{\infty}$ is observed to be independent of the lattice spacing, as found in Bogoliubov theory  Eq.~\eqref{eq:Omega_infty}.

\subsection{\label{sec:Num_method}Method to extract $\Omega_{\infty}$}

\begin{figure}[h!]
\includegraphics[width=7cm]{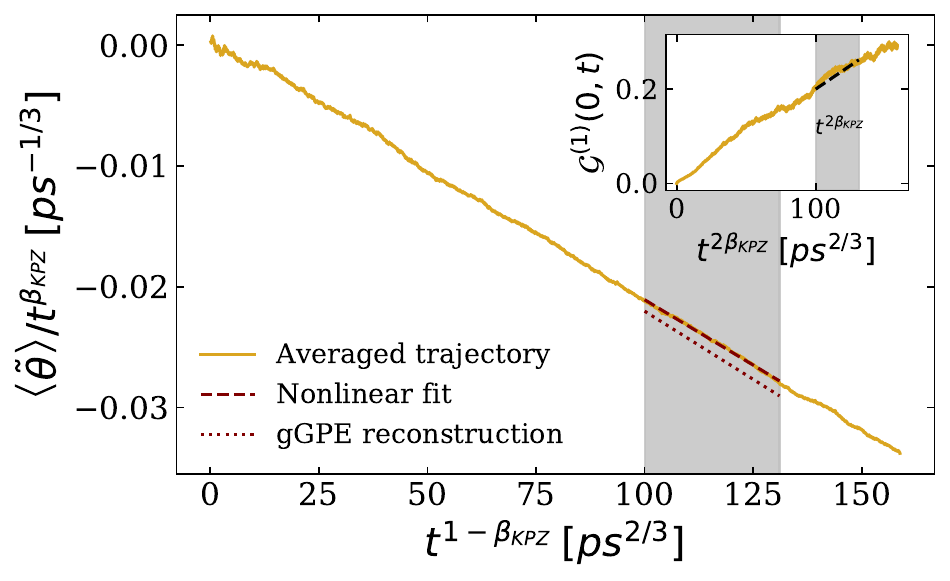}
\caption{\label{fig:Methods} Determination of $\Omega_{\infty}$, corresponding to the slope of the average  phase trajectory $\langle\Tilde{\theta}\rangle$ in the time window where the KPZ scaling is observed in time, for $\sigma=0.1\sigma_N$. This window is determined from the $\mathcal{G}^{(1)}(0,t)$ function, shown in inset (shaded area $\left[ 10^3 ps, 1.5\times10^3ps \right]$). The result from the non-linear fit and from the reconstruction of Eq.~\eqref{eq:phase_dynamics} are shown in dashed and dotted lines respectively.}
\end{figure}

In order to extract  $\Omega_{\infty}$, we use the following procedure.
The condensate dynamics is solved numerically until the BEC reaches its non-equilibrium steady state at time $t_0$, from which the first order correlation function $g^{(1)}$ and phase fluctuations trajectories defined as $\Tilde{\theta}(t+t_0) = \left[\theta(0,t+t_0) - \theta(0,t_0) \right]-\left[\theta_0(t+t_0) - \theta_0(t_0) \right]$, with $\theta_0(t)$ the mean-field expression, are recorded. The coherence $g^{(1)}$ is used to check for the presence of the KPZ regime and determine the relevant length and time scales in which it develops. The blueshift correction $\Omega_{\infty}$ is extracted in the time window in which the KPZ scaling is observed (see inset Fig.~\ref{fig:Methods}). This can be done in two ways. The first method is to perform  a non-linear fit of the average of the phase trajectories, using the asymptotic behavior \eqref{eq:omegainfanz} as fitting ansatz. A second method is to reconstruct $\Omega_{\infty} = -\langle \partial_t\Tilde{\theta} \rangle$ from the dynamical equation of the phase Eq.~\eqref{eq:phase_dynamics}, by computing each of its contributions separately. An example of numerical measurement of $\Omega_{\infty}$ following both of these procedures is shown on Fig.~\ref{fig:Methods}. We verified that the value of $\Omega_\infty$ extracted from the non-linear fit or from the reconstruction of the gGPE coincides within numerical precision in all cases. This consistency check being fulfilled, we only show data coming from the reconstruction of the gGPE in the following. The result can then be compared with the effective KPZ phase dynamics Eq.~\eqref{eq:KPZmapping}, in order to assess the
importance of the different terms neglected in the mapping.
Note that $\langle \cdot \rangle$  represents here an ensemble average over numerical realizations of the noise, discarding any phase trajectories containing defects. We checked that these defects rarely occur for the chosen parameters. In addition, we verified that the adiabatic approximation for the reservoir density holds in the parameter regime considered. In particular,  the relative error committed by estimating the average fluctuation of the reservoir density as $\langle\Tilde{n}_R\rangle=-2\dfrac{g_{i}}{R}\langle\Tilde{n}\rangle$ is small, such that the condensate depletion contribution to $\Omega_{\infty}$ can be estimated as $\langle g\Tilde{n} + 2g_R\Tilde{n}_R \rangle\approx -g_{e}\langle \Tilde{n}\rangle_S$, as written in Eq.~\eqref{eq:Omega_infty_CAB}.

\subsection{\label{sec:Num_fctsigma}Blueshift corrections as a function of the noise strength}

\begin{figure}[h!]
\includegraphics[width=7.5cm]{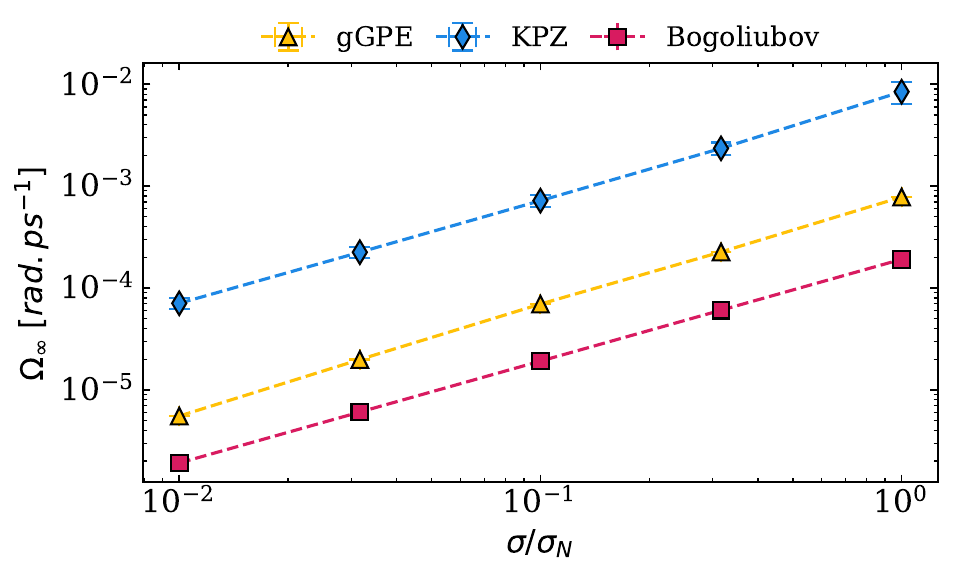}
\caption{\label{fig:blueshiftfctsigma} $\Omega_{\infty}$ as a function of the noise strength $\sigma$ extracted from the  reconstruction of the full dynamics Eq.~\eqref{eq:phase_dynamics} (orange triangle), of the effective  KPZ dynamics Eq.~\eqref{eq:KPZmapping} (blue diamonds), and of the Bogoliubov prediction Eq.~\eqref{eq:Omega_infty} (purple squares).}
\end{figure}

\begin{figure}[h!]
\includegraphics[width=7cm]{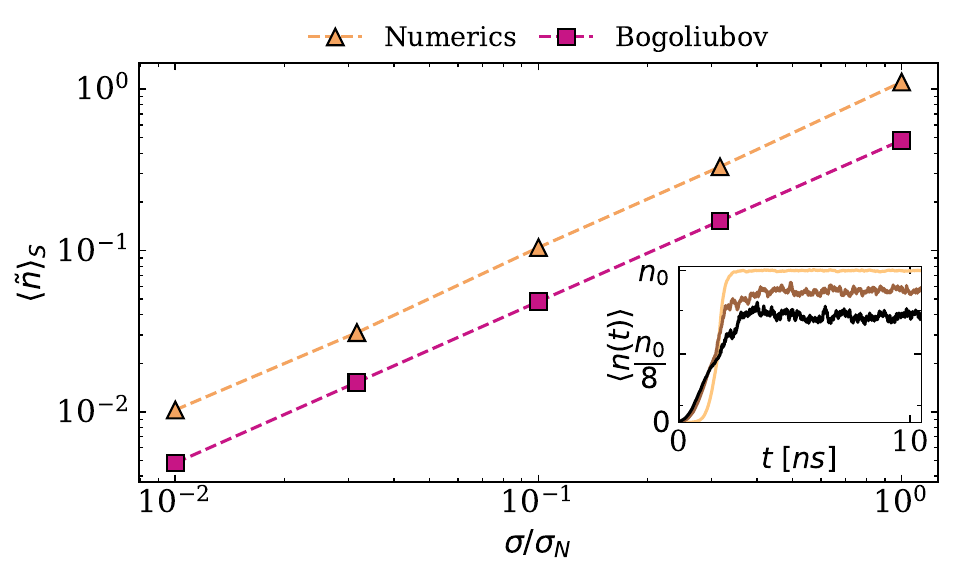}
\caption{\label{fig:condensate_depletion_sigma} Stochastic depletion $\langle\Tilde{n}\rangle_S$ as a function of the noise strength $\sigma$ from numerical resolution of the full dynamics Eq.~\eqref{eq:gGPE} (orange triangle) and the Bogoliubov prediction Eq.~\eqref{eq:stochastic_depletion} (purple squares). Inset: average density as a function of time for increasing noise (from light to dark $\sigma/\sigma_N=\{10^{-2}, 0.5, 1\}$), where the vertical scale is logarithmic from $0$ to $n_0/8$ and linear above. The stationary density $\langle n(t) \rangle$ coincides with  the mean field density $n_0$ for vanishing noise.}
\end{figure}

The activation of phase defects is exponentially sensitive to the noise strength $\sigma$ \cite{Diehlspacetimevortex, vercesi2023phase}. In order to study the dependence on the noise of the blueshift correction while maintaining  a defect-free condensate, $\sigma$ is varied from $10^{-2}\sigma_N$ to $\sigma_N$  with the integration time step $dt=0.076ps$. The results are shown in Fig~\ref{fig:blueshiftfctsigma}, and compared  with Bogoliubov theory Eq.~\eqref{eq:Omega_infty}.
We find that the numerical estimate for the blueshift correction exhibits a proportionality in the noise strength $\sigma$ consistent with the prediction \eqref{eq:Omega_infty}, with $\Omega_{\infty}\approx5\times10^{-3}\Omega_0$ in typical experimental conditions. It is however larger than the Bogoliubov prediction \eqref{eq:Omega_infty} by an approximate $3.5$ factor. This difference cannot be explained by finite size effects on the asymptotic growth velocity of the interface \cite{Krugasymptoticvelocity}, as we checked that $\Omega_{\infty}$ is already converged for condensates of length $100\mu m$. Examining the values of the terms entering in the numerical reconstruction from Eq.~\eqref{eq:phase_dynamics}, we find that the deviation from the theory is instead traced back to density fluctuations, whose contribution dominates the blueshift correction $g_{e}\langle \tilde{n} \rangle_S\approx0.7\times\Omega_{\infty}$. 
We show on Fig.~\ref{fig:condensate_depletion_sigma} the condensate depletion from both numerical simulation and the Bogoliubov calculation Eq.~\eqref{eq:stochastic_depletion}. Even though both follow the same trend, the former is larger than the latter by a factor $\approx 2$, hence explaining the result of Fig.~\ref{fig:blueshiftfctsigma}.

Moreover, we also compare in Fig.~\ref{fig:blueshiftfctsigma} the  value of $\Omega_\infty$ extracted from the full phase dynamics Eq.~\eqref{eq:phase_dynamics} and from the effective KPZ phase dynamics Eq.~\eqref{eq:KPZmapping}. Although this second estimate is proportional to $\sigma$ as well, its actual values largely differ from both gGPE reconstruction and Eq.~\eqref{eq:Omega_infty}. This difference originates from the diverging contribution to $\langle(\partial_x\Tilde{\theta})^2\rangle$ -- regularized by $dx$ in the numerics -- (see Eq.~\eqref{eq:derivative_Cthetatheta_bogo} and Fig.~\ref{fig:short_scale_rho_theta}), which is removed when considering all the other contributions. This numerically confirms the need of including both density and density-phase terms to correctly describe blueshift corrections.

\subsection{\label{sec:Num_lambda}Influence of the interaction strength on blueshift corrections}

\begin{figure}[h!]
\includegraphics[width=7cm]{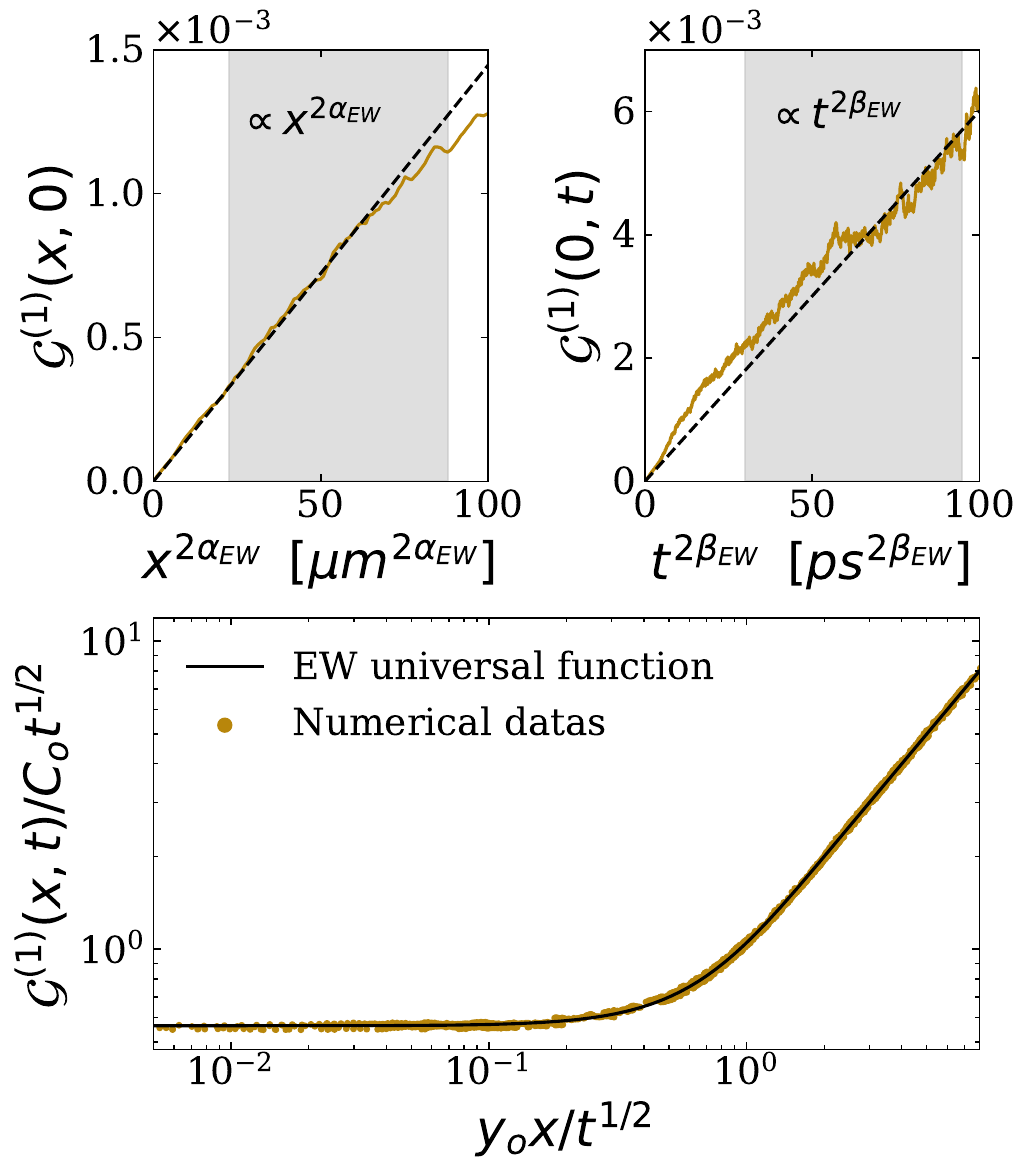}
\caption{\label{fig:EWscaling} EW scaling regime reached when $g = g_{\rm max}$: (upper left panel) spatial scaling of $\mathcal{G}^{(1)}(x,0)$ as a function of $x$, indicated by the grey shade in between $\left[22 \mu m, 88 \mu m \right]$; (upper right panel) temporal scaling of $\mathcal{G}^{(1)}(0,t)$ as a function of $t^{1/2}$, indicated by the grey shade in between $\left[9\times10^2 ps, 9\times10^3 ps \right]$; (lower panel) collpase of $\mathcal{G}^{(1)}(x,t)$ onto the EW scaling universal function $g_{\rm EW}$.}
\end{figure}

In this section, the polariton-polariton interaction strength $g$ is varied from 0 to  $g_{\rm max}\simeq 2.6\times g_R$ in order to vary  the KPZ non-linearity. Note that this also induces a change of the other KPZ parameters $\nu$, $D$, and of the length $\kappa_h$, such that  the effect of the KPZ non-linearity on $\Omega_\infty$ cannot be isolated by varying the gGPE microscopic parameters. The maximal value of $g$ studied is chosen such that
 $\lambda$ vanishes,  and the Edwards-Wilkinson limit of the effective phase dynamics \eqref{eq:KPZmapping} is reached. The noise strength is fixed to $\sigma=0.1\sigma_N$ to avoid any defect and, as the interaction strength increases, the integration time step is gradually decreased from $dt=0.076ps$ to $dt=0.010ps$ to keep the phase dynamics well resolved in time. Note that $g_{e}<0$ for all these values of $g$, such that the condensate does not exhibit any modulation instability. When the KPZ non-linearity $\lambda$ vanishes for $g=g_{\rm max}$, the scaling form of $\mathcal{G}^{(1)}(x,t)$ changes from $C^{\rm KPZ}_{hh}$ Eq.~\eqref{eq:scaling_form_G1_KPZ} to $C^{\rm EW}_{hh}$ Eq.~\eqref{eq:scaling_form_G1_EW}. We show in Fig.~\ref{fig:EWscaling} that the EW scaling regime is indeed reached when $g=g_{\rm max}$. For this value, as the KPZ scaling is not visible anymore, $\Omega_{\infty}$ is computed within the temporal window of the EW scaling.

\begin{figure}[h!]
\includegraphics[width=8cm]{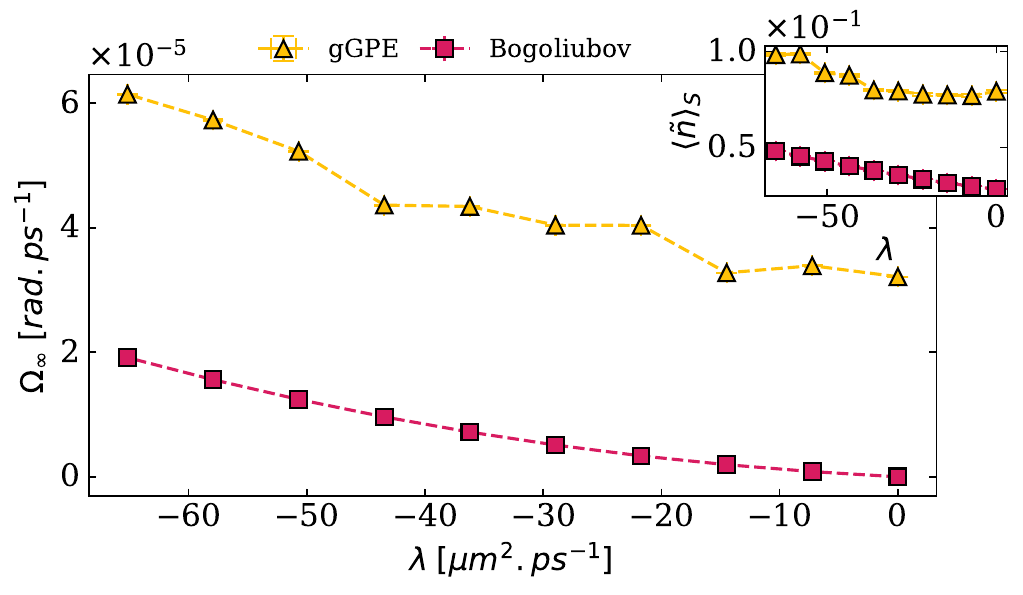}
\caption{\label{fig:blueshiftfctg} $\Omega_{\infty}$ as a function of the KPZ non-linearity $\lambda$ extracted from the  reconstruction of the full dynamics Eq.~\eqref{eq:phase_dynamics} (orange triangle) and from the Bogoliubov prediction Eq.~\eqref{eq:Omega_infty} (purple squares). Inset: condensate depletion $\langle \Tilde{n} \rangle_S$ as a function of $\lambda$.}
\end{figure}

According to the Bogoliubov prediction \eqref{eq:Omega_infty}, the blueshift correction  $\Omega_{\infty}$ should vary linearly with $\lambda$ until it vanishes in the EW regime. However, since varying $g$ also induces a variation of $\nu$ and $D$, the global behavior of \eqref{eq:Omega_infty} is more complex. We show in Fig.~\ref{fig:blueshiftfctg} the variation of the blueshift correction computed both from the Bogoliubov prediction \eqref{eq:Omega_infty} and from  reconstruction of the full dynamics Eq.~\eqref{eq:phase_dynamics}, as a function of the effective KPZ non-linearity $\lambda$. We observe that it decreases with $\lambda$,  in an algebraic way not too far from linear. This indicates that the variation of $\lambda$ is a dominant contribution to the behavior of $\Omega_{\infty}$ when varying the interaction strength $g$. Moreover, for the Bogoliubov prediction, it  vanishes as expected when $\lambda=0$ and the EW limit is reached.
The blueshift correction $\Omega_{\infty}$ evaluated from the full  gGPE qualitatively exhibits the same decay with $\lambda$ as predicted from Eq.~\eqref{eq:Omega_infty}, but it does not vanish at $\lambda=0$. As in Sec.~\ref{sec:Num_fctsigma}, the shift between these two curves can be explained by the discrepancy between the predicted and numerically observed condensate depletion (shown in the inset of Fig.~\ref{fig:blueshiftfctg}).

\subsection{{ Experimental observability}}

{ We restricted so far our theoretical analysis to idealized parameter conditions, in particular small noise levels, in order to damp the emergence of defects in the phase. Under these conditions, the overall stochastic correction to the blueshift $\Omega_\infty$ is found to be very small compared to the mean-field contribution $\Omega_0$, typically of order $\Omega_\infty/\Omega_0\simeq 5\times 10^{-3}$, which makes it hardly observables in actual experiments. From the Bogoliubov expression~\eqref{eq:Omega_infty}, this correction increases linearly with the noise strength, and also with $g_R$ as $\sim g_R^{3/2}$.}

{ The noise level can be increased in experiments in a controllable fashion by increasing the temperature of the cryostat. This  leads to a growth of the population of the phonons of the semiconductor material. Such phonons are coupled to the cavity polaritons via the exciton-phonon coupling, and act as an additional bath for the polaritons \cite{Shelykh, anna-maxime}.  
We performed additional simulations up to twice the nominal noise. In this regime, we found that the KPZ universal properties are resilient to the presence of some defects and for $\sigma=2\sigma_0$ we obtain $\Omega_\infty/\Omega_0\simeq 10^{-2}$, which becomes experimentally detectable. For higher values of the noise strength, the number of phase defects rapidly increases as shown in \cite{Diehlspacetimevortex,vercesi2023phase}, making it difficult to distinguish between the linear evolution of the average unwrapped phase caused by blueshift corrections $\Omega_{\infty}$ and the evolution arising from phase defects.}

{ The interaction strength with the reservoir $g_R$ can be also experimentally controlled to some extent  by increasing the number of quantum wells in the optical microcavities. However, in our parameter setting, increasing $g_R$ readily leads to modulation instabilities for which the condensate exhibits many phase defects.
It should be emphasized that the estimated values for $\Omega_{\infty}$ as well as the appearance of modulation instability depend on the whole set of parameters of the model. In our simulations, we have  chosen parameters close to the experiment of Ref.~\cite{Bloch&co2021}, but
by changing the design of the experimental sample and conditions, one could explore a large parameter space, which may allow one to  
increase blueshift corrections without triggering the modulational instability.
}

\section{Conclusions}
\label{sec:conclusions}

We showed that the blueshift of a 1D exciton-polariton condensate fluctuates around its mean-field value, resulting in a non-zero average correction $\Omega_\infty$. This correction  originates from the non-linear stochastic dynamics of the driven-dissipative condensate and was studied in its defect-free KPZ phase. Using Bogoliubov theory, we determined an analytical expression for  $\Omega_\infty$, which, although sensitive to  short-distance properties of the condensate correlation functions, can surprisingly be summarized in terms of both the KPZ effective parameters and  characteristic length scales of the condensate.
We then performed numerical simulations of the full dynamics, varying both the noise and the interaction strength, to check this prediction.  We highlighted in particular that spatial variations of the density, which can be safely neglected to study the long-distance properties of the phase, give in contrast an important contribution  to the blueshift correction.

{ With idealized parameter conditions, in particular small noise levels, the overall stochastic correction to the blueshift is very small.  However, we verified that our predictions remain valid for larger noise strength up to twice the nominal noise, for which it could become detectable. As a consequence, blueshift corrections  could in principle be accessed in experimental platforms where phonon-polariton interaction strength is sufficiently strong.}
{In outlook, it would be interesting to investigate how to enhance this correction in other ways,  for example exploring other regions of the parameter space, in order to make it more easily observable in experiments. }



\begin{acknowledgments}
FH acknowledges support from the Laboratoire d’excellence LANEF in Grenoble (ANR-10-LABX-51-01). AM acknowledges funding from the Quantum-SOPHA ANR Project ANR-21-CE47-0009. LC acknowledges support  from Institut Universitaire de France (IUF).
\end{acknowledgments}

\appendix

\section{\label{sec:ap_bogo}Bogoliubov theory}

\noindent
We start from the dynamical equations for the condensate and reservoir
Eqs.~\eqref{eq:gGPE} and \eqref{eq:xreservoir}, setting $\hbar = 1$ for simplicity of notations:
\begin{eqnarray}
i\partial_t\Phi & = & \Bigg[ \mathcal{F}^{-1}\left[ \epsilon_{\hat{k}} - \dfrac{i}{2}\gamma_{\ell}(\hat{k}) \right] + \dfrac{i R}{2}n_R \nonumber\\
    & & +  g|\Phi|^2 + 2 g_Rn_R  \Bigg]\Phi + \xi \label{eq:ap_gGPE} \\
\partial_tn_R & = & P - \left(\gamma_R + R|\Phi|^2 \right)n_R \label{eq:ap_xreservoir}\,,
\end{eqnarray}
where $\langle \xi \rangle=0$ and $\langle \xi(x,t)\xi^*(x',t') \rangle=2\sigma\delta(x-x')\delta(t-t')$. In the adiabatic approximation of the reservoir $\partial_tn_R\approx 0$, Eq.~\eqref{eq:ap_xreservoir} reduces to $n_R = P/(\gamma_R+R|\Phi|^2)$. Expanding the wavefunction around its fluctuations $\Phi = e^{-i\Omega_0t}\left[ \sqrt{n_0} + \delta\Phi(x,t)\right]$ and injecting it into Eq.~\eqref{eq:ap_gGPE} leads at zeroth order to $n_{R0}=\gamma_0/R$ and $\Omega_0 = gn_0 + 2g_Rn_{R0}$. At first order in $\delta\Phi$, Eq.~\eqref{eq:ap_gGPE} yields the linear dynamics
\begin{eqnarray}
    i\partial_t\begin{pmatrix}
    \delta\Phi \\
    \delta\Phi^*
    \end{pmatrix} = \mathcal{F}^{-1}\left[\mathcal{L}\right]\begin{pmatrix}
    \delta\Phi \\
    \delta\Phi^*
    \end{pmatrix} + \sqrt{\sigma}\begin{pmatrix}
\Tilde{\xi} \\
-\Tilde{\xi}
\end{pmatrix},
\label{eq:ap_Bogo_phi}
\end{eqnarray}
where we introduced $\Tilde{\xi} = \xi e^{-i\Omega_0t}$ and the Bogoliubov matrix
\begin{eqnarray}
    \mathcal{L}=
    \begin{pmatrix}
        \epsilon_{\hat{k}} + g_{e}n_0 + i\Gamma_{\hat{k}} & g_{e}n_0 - ig_{i}n_0 \\
        - g_{e}n_0 - ig_{i}n_0 & -\left( \epsilon_{\hat{k}} + g_{e}n_0  \right) - i\Gamma_{\hat{k}}
    \end{pmatrix},
    \label{eq:ap_Bogomatrix_L}
\end{eqnarray}
with $\Gamma_{\hat{k}} = \frac{\Delta\gamma(\hat{k})}{2} - g_{i}n_0$, $\Delta\gamma(\hat{k}) = Rn_{R0}-\gamma_{\ell}(\hat{k})$ and $g_{i}=(Rn_{R0})^2/2P$. 

We introduce the rotation matrix 
\begin{equation}
    \mathcal{A} = \dfrac{1}{2\sqrt{n_0}}
    \begin{pmatrix}
    1 & 1 \\
    -i & i
    \end{pmatrix}
    \label{eq:ap_A}
\end{equation} 
such that
\begin{eqnarray}
    \begin{pmatrix}
    \delta n/2n_0 \\
    \delta\theta
    \end{pmatrix} = \mathcal{A}
    \begin{pmatrix}
    \delta\Phi \\
    \delta\Phi^*
    \end{pmatrix}\,.
\label{eq:ap_transfo_phi_n_theta}
\end{eqnarray}
The   dynamics of the density and phase   is given by 
\begin{eqnarray}
    \left(\partial_t - \mathcal{F}^{-1}\left[\mathcal{L}_{rot}\right]\right)\begin{pmatrix}
    \delta n/2n_0 \\
    \delta\theta
    \end{pmatrix} =\sqrt{\dfrac{\sigma}{n_0}}\begin{pmatrix}
{\rm Im}\left(\Tilde{\xi}\right) \\
-{\rm Re}\left(\Tilde{\xi}\right)
\end{pmatrix}
\label{eq:ap_Bogo_n_theta}
\end{eqnarray}
with
\begin{equation}
    \mathcal{L}_{rot} = \begin{pmatrix}
        \Gamma_{\hat{k}}-g_{i}n_0   &  \epsilon_{\hat{k}}  \\
        -\left( \epsilon_{\hat{k}} + 2g_{e}n_0 \right) & \Gamma_{\hat{k}}+g_{i}n_0 
    \end{pmatrix}
    \label{eq:ap_Lrot}
\end{equation}
whose eigenvalues read 
\begin{equation}
    \omega_{\pm} = -i\Gamma_{\hat{k}} \pm \sqrt{\epsilon_{\hat{k}}\left( \epsilon_{\hat{k}} + 2g_{e}n_0\right) - g_{i}^2n_0^2}.
    \label{ap:bogoEigenval}
\end{equation}
From Eq.~\eqref{eq:ap_Bogo_n_theta}, and after integrating over the frequency $\omega$, we obtain the equal time correlation functions

\begin{eqnarray}
    \langle \delta n(k) \delta n(-k) \rangle & = & \dfrac{-\sigma n_0}{\Gamma_{\hat{k}}}\times \dfrac{ \epsilon_{\hat{k}}^2 + \Delta \gamma(\hat{k})^2/4 }{E_{\hat{k}}^2 + \Gamma_{\hat{k}}^2 - g_{i}^2n_0^2} \nonumber \\
    && -\dfrac{\sigma n_0}{\Gamma_{\hat{k}}} \label{eq:nn_k_general} \\
    \langle \delta \theta(k) \delta \theta(-k) \rangle & = & \dfrac{\langle \delta n(k)\delta n(-k) \rangle}{4n_0^2} + \dfrac{\sigma}{\Gamma_{\hat{k}}n_0} \nonumber \\
    && \times \dfrac{\Gamma_{\hat{k}} g_{i}n_0 - \epsilon_{\hat{k}}g_{e} - g_{e}^2n_0^2 }{E_{\hat{k}}^2 + \Gamma_{\hat{k}}^2 - g_{i}^2n_0^2}  \label{eq:thetatheta_k_general}\\
    \langle \delta n(k) \delta \theta(-k) \rangle & = & - \dfrac{\sigma n_0}{\Gamma_{\hat{k}}}\times \dfrac{ g_{i}\epsilon_{\hat{k}} + g_{e}\Delta\gamma(\hat{k}) }{E_{\hat{k}}^2 + \Gamma_{\hat{k}}^2 - g_{i}^2n_0^2} \label{eq:ntheta_k_general}
\end{eqnarray}
where $E_{\hat{k}}=\sqrt{\epsilon_{\hat{k}}\left(  \epsilon_{\hat{k}} + 2g_{e}n_0\right)}$.

\section{\label{ap:k-dependent_dissipation}Effect of the $k$-dependent dissipation}

The momentum dependence of the loss rate $\gamma_{\ell}(\hat{k})$ affects the large-momentum behavior of the Bogoliubov eigenvalues Eq.~\eqref{ap:bogoEigenval} through the effective loss rate $\Gamma_{\hat{k}}$, similarly to the frequency dependent amplification of Ref.~\cite{chiocchetta2013}. This is illustrated on Fig.~\ref{fig:Bogo_spec_effect_gamma_K} below, where the Bogoliubov eigenvalues are shown for three experimentally relevant loss rates.
\begin{figure}[h!]
\includegraphics[width=7cm]{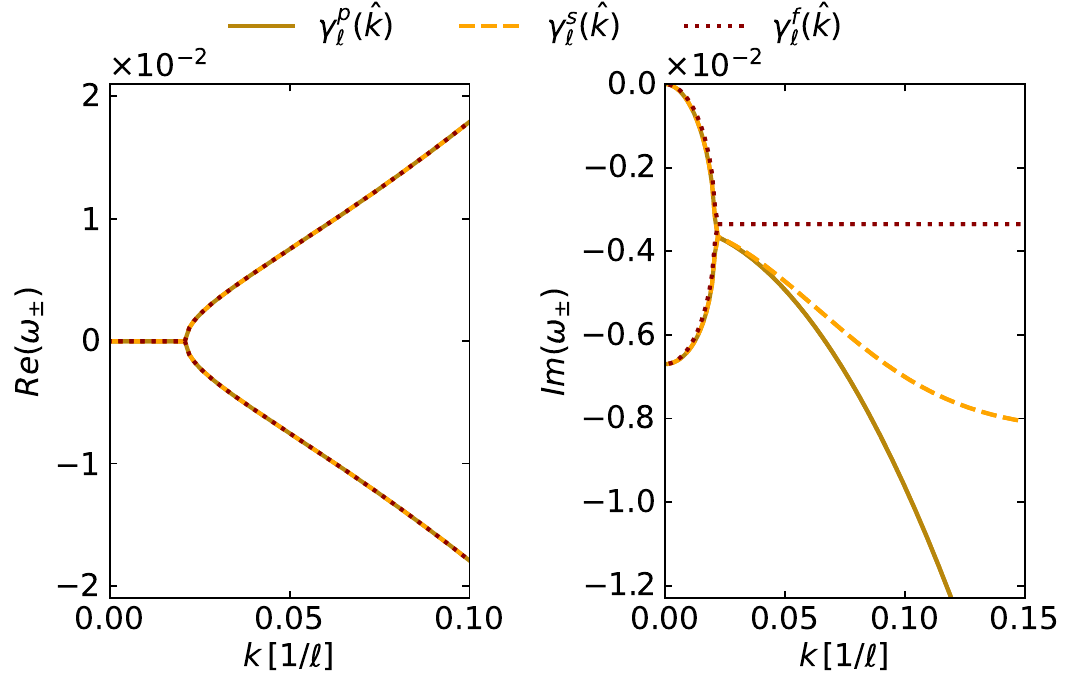}
\caption{\label{fig:Bogo_spec_effect_gamma_K} Real (left panel) and imaginary (right panel) parts of the Bogoliubov spectrum Eq.~\eqref{ap:bogoEigenval} for different losses $\gamma_{\ell}(\hat{k})$. Red dotted line: flat dissipation $\gamma_{\ell}^f(\hat{k}) = \gamma_0$. Brown plain line: parabolic dissipation $\gamma_{\ell}^p(\hat{k}) = \gamma_0 + \gamma_2k^2$. Orange dashed line: saturated dissipation $\gamma_{\ell}^s(\hat{k}) $ defined in Eq.~\eqref{eq:saturated_loss_rate}, with $\gamma_{sat}=55\mu eV$.}
\end{figure}
While the momentum independent loss rate $\gamma_{\ell}^f(\hat{k}) = \gamma_0$ is most of the time sufficient to describe the phenomenology of EP BECs, it is rather observed that the exciton-polariton linewidth broadens quadratically at small $k$ before saturating to $\gamma_{sat}$ at large momentum, which is modelled by the following saturated loss rate
\begin{eqnarray}
    \gamma_{\ell}^s(\hat{k}) = \gamma_0\dfrac{1+\alpha}{e^{-\beta k^2}+\alpha},
    \label{eq:saturated_loss_rate}
\end{eqnarray}
where $\alpha = \frac{\gamma_0}{\gamma_{\rm sat}-\gamma_0}$ and $\beta = \frac{\gamma_2}{\gamma_0}(1+\alpha)$ \cite{Bloch&co2021}. Expanding \eqref{eq:saturated_loss_rate} at small $k$, the polariton linewidth can be approximated by a parabola $\gamma_{\ell}^s(\hat{k})\approx \gamma_{\ell}^p(\hat{k})=\gamma_0 + \gamma_2k^2$.

Let us now discuss the consequences of the shape of the loss rate on blueshift corrections. Replacing $\epsilon_{\hat{k}} = \hbar^2\boldsymbol{k}^2/2m$ and $\gamma_{\ell}^f(\hat{k})$ in Eqs.~\eqref{eq:nn_k_general}-\eqref{eq:ntheta_k_general}, we recover the equal time correlations given in \cite{chiocchetta2013}. Inserting them in Eq.~\eqref{eq:Omega_infty_CAB_lambda} we find that their UV diverging contributions (discussed in the main text and Ref.~\cite{chiocchetta2013}) compensate such that $\Omega_{\infty}=\kappa_h D$ remains well defined. This leads to the estimate $\Omega_{\infty}\approx \Omega_{0}$/10 for $\gamma_{\ell}^f(\hat{k})$. A similar value is expected from the saturated linewidth $\gamma_{\ell}^s(\hat{k})$. In this work, as in experimental realizations of EP condensates, the pump $P$ extends over a large number of sites. As a consequence, the exciton reservoir, hence the condensate, are occupied in the vicinity of the momentum $k=0$ only. This implies that the parabolic linewidth $\gamma_{\ell}^p(\hat{k})$ is the most relevant to describe losses in the condensate.

\section{\label{ap:higher_dim}Blueshift corrections in higher dimensions}

The Bogoliubov approximation developed in Sec.~\ref{sec:Bogo} and Appendix~\ref{sec:ap_bogo} can be extended to higher dimensions by  replacing $\partial_x\to\nabla$ and $k\to\boldsymbol{k}$. 
Note that for $\epsilon_{\hat{k}} = \hbar^2\boldsymbol{k}^2/2m$ and independently of the specific choice of $\gamma_{\ell}(\hat{\boldsymbol{k}})$, the connected phase correlation function $C_{\theta\theta}(\boldsymbol{r})$ contains a term $\propto \int \dfrac{d\boldsymbol{k}}{\left(2\pi\right)^{d}}\dfrac{e^{i\boldsymbol{k}\cdot\boldsymbol{r}}-1}{k^2}$. This term yields a Edwards-Wilkinson contribution $C^{\rm EW}(\boldsymbol{r},0)$ to the phase spatial correlations, of the form $C^{\rm EW}(\boldsymbol{r},0)\sim|r|^{2\alpha}$  where $\alpha = \frac{2-d}{2}$ is the EW spatial exponent. { Similarly in time, one would find $C^{\rm EW}(0,t)\sim t^{2\beta}$ with $\beta = \frac{2-d}{4}$. In a one dimensional system, this gives a stretched exponential decay of the first order correlation function $|g^{(1)}(r,t)|^2\propto e^{-\sqrt{t}}$ followed for a finite size system by Schawlow-Townes exponential decay $|g^{(1)}(r,t)|^2\propto e^{-t}$ \cite{amelio2020theory}. As commented in the main text, a similar exponential decay is obtained for equilibrium condensates at finite temperature \cite{giamarchi-book}.}

In addition, the UV divergences mentioned in Sec.~\ref{sec:Bogo} for $\gamma_{\ell}(\hat{\boldsymbol{k}})=\gamma_0 + \gamma_2\hat{\boldsymbol{k}}^2$ become more severe when increasing the spatial dimension.  Bogoliubov theory remains valid provided $\langle (\nabla \delta \theta)^2 \rangle\propto\frac{\sigma}{n_0\gamma_2\ell^d}$ is a small parameter. In fact, if we first set $\gamma_2= 0$ and perform again the calculation, we recover the same condition for the validity of Bogoliubov theory as in Ref.~\cite{Castin2003extension}, namely that $\langle (\ell \nabla \delta \theta)^2\rangle\propto \frac{\sigma}{n_0\ell^d}$ is a small parameter, where the driven-dissipative nature of the condensate explicitly appears through the noise strength $\sigma$.
Nevertheless, $\Omega_{\infty}$ remains well-defined in the UV regime for $d\leq 3$ and follows the general expression
\begin{eqnarray}
    \Omega_{\infty,d} & = & -\dfrac{\lambda D}{2\nu} \kappa_h^2\kappa_d^2 \int \dfrac{ d\boldsymbol{k}/(2\pi)^d }{\left( k^2+\kappa_h^2 \right)\left( k^2+\kappa_d^2 \right)}.
    \label{eq:ap_higher_d_Omega_infty}
\end{eqnarray}
Setting $d=1$ in Eq.~\eqref{eq:ap_higher_d_Omega_infty}, we recover Eq.~\eqref{eq:Omega_infty} of the main text.

In $d>1$, the linearized theory cannot capture any effect from the KPZ physics. This is because the KPZ fixed point becomes genuinely non-perturbative in $d>1$ \cite{Canet2010}. It was shown that the perturbation theory in $\lambda$ fails to all orders to access the KPZ fixed point \cite{Wiese98}. Thus, the results obtained from Bogoliubov theory in $d>1$ cannot be related to KPZ physics. We still express them for convenience in terms of the KPZ parameters, although this should not be interpreted as resulting from non-linear KPZ physics. 

For $d=2$ we find by parity that $\Omega_{\infty}^{d=2} = 0$, thus the stochastic dynamics gives no correction to the blueshift within the Bogoliubov approximation.
The case $d=3$ is of fundamental interest for EP BECs, as the driven-dissipative condensate is predicted to recover some characteristics of its equilibrium counterpart \cite{siebererdiehl2023universality}. For example, it was shown in Ref.~\cite{chiocchetta2013} within Bogoliubov approximation that the IR divergence of the momentum distribution, which is a manifestation of the Mermin-Wagner-Hohenberg theorem for $d<2$, is suppressed for $d=3$. Setting $d=3$ in Eq.~\eqref{eq:ap_higher_d_Omega_infty}, we find $\Omega_{\infty}^{d=3} = \dfrac{\lambda D}{8\pi\nu}\dfrac{\kappa_h\kappa_d}{\kappa_h^{-1} + \kappa_d^{-1}}$. In typical parameter regimes, this blueshift correction is approximately 500 times smaller than for $d=1$. This is consistent with the idea that fluctuations are enhanced in low dimensional systems.

\bibliographystyle{prsty}
\providecommand{\noopsort}[1]{}\providecommand{\singleletter}[1]{#1}%
%

\end{document}